\documentstyle[11pt,psfig]{article}
\begin{document}
\renewcommand{\thepage}{ }
\begin{titlepage}
\title{
\hfill
\vspace{1.5cm}
{\center \bf Coarsening on percolation clusters:
out-of-equilibrium dynamics versus non linear response
}
}
\author{
P. Butaud\thanks{E-mail: butaud@crtbt.polycnrs-gre.fr}
and R. M\'elin\thanks{E-mail: melin@crtbt11.polycnrs-gre.fr}\\
{}\\
{Centre de Recherches sur les Tr\`es basses temp\'eratures (CRTBT-CNRS)}\\
{Laboratoire conventionn\'e avec l'Universit\'e Joseph Fourier}\\
{BP 166X, 38042 Grenoble C\'edex, France}\\
}
\date{}
\maketitle
\begin{abstract}
\normalsize
We analyze the violations of linear fluctuation-dissipation
theorem (FDT) in the coarsening dynamics
of the antiferromagnetic Ising model
on percolation clusters
in two dimensions. The equilibrium magnetic response is
shown to be non linear for magnetic fields of the
order of the inverse square root of the number of sites.
Two extreme regimes can be identified in the thermoremanent
magnetization:
(i) linear response and out-of-equilibrium relaxation
for small waiting times (ii) non linear response
and equilibrium relaxation for large waiting times.
The function $X(C)$ characterizing the
deviations from linear FDT cross-overs from unity
at short times to a finite positive value for
longer times, with the same qualitative behavior
whatever the waiting time. We show that the coarsening
dynamics on percolation clusters exhibits stronger
long-term memory than usual euclidian coarsening.
\end{abstract}
\end{titlepage}

\newpage
\renewcommand{\thepage}{\arabic{page}}
\setcounter{page}{1}
\baselineskip=17pt plus 0.2pt minus 0.1pt
\section{Introduction}
Aging experiments in spin glasses
\cite{Lundgren,Saclay}
first carried out by Lundgren {\it et al.},
have generated a large amount of experimental
as well as theoretical work. Two types
of experiments have been investigated: the
zero-field-cooled experiment and the
thermoremanent magnetization experiment,
both leading to similar results. In the present
article, we will restrict ourselves to the
thermoremanent magnetization experiment,
consisting in first quenching the system below
its glass transition temperature at time
$t=0$, applying a small constant magnetic field up
to the waiting time $t_w$, switching off the magnetic
field at time $t_w$, and measuring the magnetization
relaxation at time $t_w + \tau$. It is an
experimental observation that the
magnetization relaxation depends on the
``age'' of the system, namely, on the waiting time.
Different theoretical approaches have been
developed so far, for instance: droplet
picture \cite{droplet1,droplet2}, mean field
models \cite{mean_field} or phenomenological trap
models \cite{traps}.
Several scenarios have been proposed, such as
``true'' versus ``weak'' ergodicity breaking \cite{traps},
or ``interrupted'' aging \cite{traps}. The first two
scenario depend on whether ergodicity breaking occurs
for finite or infinite waiting times. ``Interrupted''
aging means that, at a finite temperature, there
is no more aging if the waiting time is larger than  finite
(but possibly large) time scale. In other words, the system
equilibrates in a finite time.

Aging can be characterized by the violation of the
fluctuation-dissipation theorem (FDT).
If the system has reached thermodynamic equilibrium before
the magnetic field is switched off at time $t_w$,
the magnetization response is then independent on $t_w$.
This situation can be realized either at large temperatures,
or in an ``interrupted aging'' situation (which will be the case
in the present article), or, in the
presence of non-interrupted aging, by formally taking
the limit $t_w \rightarrow + \infty$ before
the thermodynamic limit $N \rightarrow + \infty$.
Then,
the equilibrium thermoremanent magnetization $m(\tau)$
is related in a simple fashion to the autocorrelation of
the spin configurations at times $t_w$ and $t_w + \tau$
via the FDT (see section \ref{sec-thermoremanent}
for more details).
In the out-of-equilibrium dynamics, the FDT
is no more valid, and there are analytical
predictions in some mean-field solvable models
of what is the FDT violation
\cite{mean_field,mean_field_2,mean_field_3}.
In particular, 
Cugliandolo and Kurchan \cite{mean_field} have
proposed that the
out-of-equilibrium linear response kernel $R(t,t')$
relating the magnetization to the correlation
depends on $t$ and $t'$ only through the autocorrelation
$C(t,t')$. The FDT violation is then characterized
by a function $X(C)$ that depends only on the
autocorrelation, and, as recalled in
section \ref{sec-thermoremanent}, can be
obtained from the thermoremanent magnetization
simulations.

The aim of the present article is to study the FDT violation
in dilute Ising antiferromagnets at the percolation threshold,
with a Hamiltonian
\begin{equation}
\label{eq-hamiltonian}
H = J \sum_{\langle i,j \rangle}
\sigma_i \sigma_j
,
\end{equation}
where the summation is carried out over neighboring
pairs of spins on a percolation cluster. In practice,
will will study here only percolation clusters
generated on a square lattice in a two dimensional
space.
It is well-known that, for a finite cluster of $N$ sites,
the dynamics freezes as the temperature is decreased below
the glass cross-over temperature \cite{Rammal-Benoit}
\begin{equation}
\label{eq-Tg}
T_g = \frac{2 J \overline{d} \nu_P}{\ln{N}}
,
\end{equation}
with $\overline{d}$ the fractal dimension and
$\nu_P$ the percolation exponent. This glass
cross-over originates from the conjugate
effect of large-scale
`droplet' excitations (with
zero temperature energy barriers that scale like
$J \ln{N}$ \cite{Rammal,Henley}),
and the divergence of the correlation length
at low temperatures \cite{Coniglio}.

It is of interest to understand the FDT
violation in these systems for two reasons.
First, a quite
different behavior from euclidian coarsening is expected,
with more  pronounced long-term memory effects 
due to the slow dynamics of `droplet' excitations.
We will
indeed show that the function $X(C)$ characterizing
the fluctuation-dissipation ratio cross-overs
from unity to a smaller value $X_0$ in the aging regime.
Whereas $X_0$ is zero in euclidian coarsening, we find a
non-zero value for coarsening on percolation clusters. This
indicates that, even though these non-frustrated systems show
interrupted aging, the FDT violation 
in these systems shares some common features with
``true'' spin-glasses.
The second motivation for studying these systems is, as will
be developed in section \ref{sec-equilibrium}, that the
low temperature magnetic response to an
external magnetic field is linear only for
magnetic fields smaller than a typical field $h^*$ scaling like
$T/\sqrt{N}$ (see section \ref{sec-equilibrium}).
One is thus lead to
study the FDT violation in the absence of linear response
to an external magnetic field.

This article is organized as follows: section
\ref{sec-equilibrium} is devoted to analyzing
the equilibrium magnetic response to an external magnetic
field and to show that the low temperature 
equilibrium response is non linear. Section
\ref{sec-thermoremanent}
recalls how the function $X(C)$ characterizing the
FDT violation can be obtained from the thermoremanent
magnetization experiment.
The results of our simulations
are next presented and discussed in section \ref{sec-results}.

\section{Absence of linear response at low temperature}
\label{sec-equilibrium}
In this section, we analyze the low temperature
equilibrium response to an external magnetic field.
We consider a percolating cluster of $N$ sites, and
first analyze a toy model for the magnetization response
to an external magnetic field.
The equilibrium
magnetization $M(h)$ in an external magnetic field $h$
can be expressed as
\begin{equation}
\overline{M(h)} =
\overline{ \frac{ \partial \ln{\langle \exp{(\beta h M)} \rangle_0}}
{\partial (\beta h)}}
\label{eq-expansion}
,
\end{equation}
where $\langle X \rangle_0$ denotes the thermal average
of the observable $X$ with respect to the system
without a magnetic field, and $\overline{X}$
denotes the disorder average. 

We are first
going to formulate in section \ref{sec-formulation}
a low-temperature toy-model which allows analytical
calculations of (\ref{eq-expansion}).
The predictions from
this toy-model will be compared to simulations
in section \ref{sec-comparison}.

\subsection{Formulation of the low temperature toy-model}
\label{sec-formulation}
Our toy-model relies on
some assumptions about the geometry of the percolating
clusters, and further assumptions about the low temperature
magnetization distributions. The validity of these
assumptions relies on the fact that some of the predictions
of our toy-model can be successfully compared
to simulations (see section \ref{sec-comparison}).

\begin{figure}
\centerline{\psfig{file=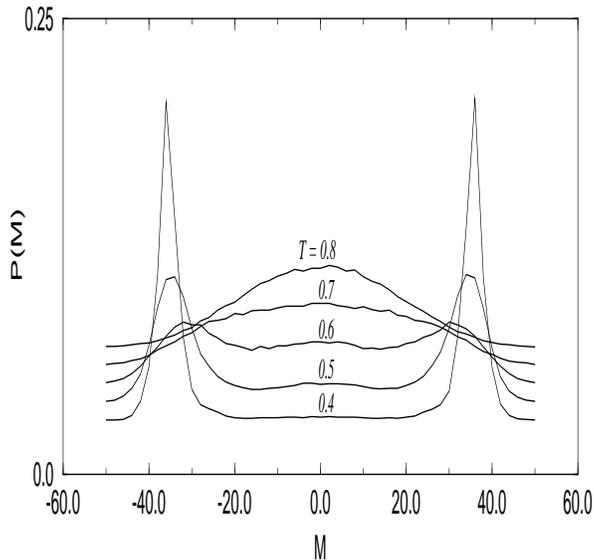,width=8cm,height=9cm}}
\caption{Magnetization distribution of a cluster
of $N=836$ sites,with $\Delta=36$. This cluster is
shown in Fig. \protect \ref{fig-dessins} (cluster A).}
\label{fig-PM}
\end{figure}

In the dilute
antiferromagnets model (\ref{eq-hamiltonian}),
the magnetization of the N\'eel state
of the percolating cluster
is equal, up to a sign, to the difference
$\Delta = N_A - N_B$ 
in the number of sites in the two sublattices A and B,
the number of sites in the percolating cluster
being $N=N_A + N_B$.
In order to allow for analytic treatment,
we assume that both $N_A$ and $N_B$ are independent
variables and gaussian distributed according to
\begin{equation}
P(N_{A,B}) = \frac{1}{\sqrt{2 \pi} \sigma}
\exp{ \left\{- \frac{1}{2 \sigma^2}
(N_{A,B} - \frac{1}{2} {\bar N})^2  \right\}}
,
\end{equation}
with a width $\sigma$ scaling like $\sigma \propto
\sqrt{\bar N}$. Within these assumptions, the
distribution of the N\'eel state magnetization
$\Delta$ is also gaussian distributed:
\begin{equation}
\label{eq-P-Delta}
P(\Delta) = \frac{1}{2 \sqrt{ \pi} \sigma}
\exp{ \left\{- \frac{1}{4 \sigma^2}
\Delta^2  \right\}}
.
\end{equation}
At zero temperature, the magnetization distribution
of a given percolation cluster consists of two delta functions
located at $M = \pm \Delta$. As shown in Fig.
\ref{fig-PM}, the effect of a small temperature
is a broadening of the two peaks at $\pm \Delta$.
The numerical calculations of $P(M)$ shown
in Fig. \ref{fig-PM} were carried
out using the Swendsen-Wang algorithm \cite{Swendsen}.

In our toy-model, we first make the approximation
that all the geometry-dependence of the magnetization
response is encoded in the single parameter $\Delta$.
This approximation becomes exact in the zero temperature
limit. At a finite but sufficiently low temperature,
we still assume this single-parameter description
of geometric fluctuations. 
We further assume that the effect of a finite temperature
is a gaussian broadening of the peaks at
$\pm \Delta$ in the magnetization distribution:
\begin{equation}
\label{eq-PM-toy}
P_{\Delta,\sigma_{\beta}}(M) = 
\frac{1} {2 \sqrt{2 \pi} \sigma_{\beta}}
\exp{ \left\{ \frac{1}{2 \sigma_{\beta}^2}
(M- \Delta)^2 \right\}} +
\frac{1} {2 \sqrt{2 \pi} \sigma_{\beta}}
\exp{ \left\{ \frac{1}{2 \sigma_{\beta}^2}
(M + \Delta)^2 \right\}}
.
\end{equation}
The thermal broadening $\sigma_{\beta}$ originates
from low temperature excitations above the
N\'eel state. At sufficiently low temperatures,
only the lowest energy excitations contribute
to $\sigma$. If the system was not diluted,
these excitations are properly described
as a dilute gaz of clusters of spins with
a wrong orientation with respect to the
N\'eel state. The contribution to
$\sigma_{\beta}$ of these excitations
is of the order of $\sqrt{N} f(T)$,
$f$ being $N$-independent and behaving like
$\ln{f(T)} \sim - J/T$. It is also
well-known that long range low
energy `droplets' also exist 
in dilute percolating antiferromagnets, due to the
self-similarity of the structure, and the
fact that the order of ramification of
the lattice is finite \cite{LesHouches}:
it costs a finite
energy to isolate a cluster of arbitrary
size from the rest of the structure.
These `droplet' excitations can
be clearly identified in the
magnetization distribution of the 
ferromagnetic Ising model \cite{moi}.
In fact, we can account for these
droplet excitations in an effective distribution
of the parameter $\sigma_{\beta}$
over the N\'eel state magnetization $\Delta$.
We will come back on this point
latter on in section \ref{sec-comparison}.

We have chosen in (\ref{eq-PM-toy})
a gaussian contribution of thermal excitations.
The resulting contribution to the magnetization
of these thermal excitations is linear in the
magnetic field. In fact, in order to describe
how the magnetization saturates for magnetic fields
scaling like $N^0$, one should refine our
toy-model to incorporate non gaussian tails
in the magnetization distribution. Since,
as mentionned previously, we are mainly interested
in magnetic fields scaling like $T/\sqrt{N}$,
these non gaussian tails do not play any significant
role in this low magnetic field physics.

\subsection{Non linear effetcs}
\label{sub-sec-non-linear}
\begin{figure}
\centerline{\psfig{file=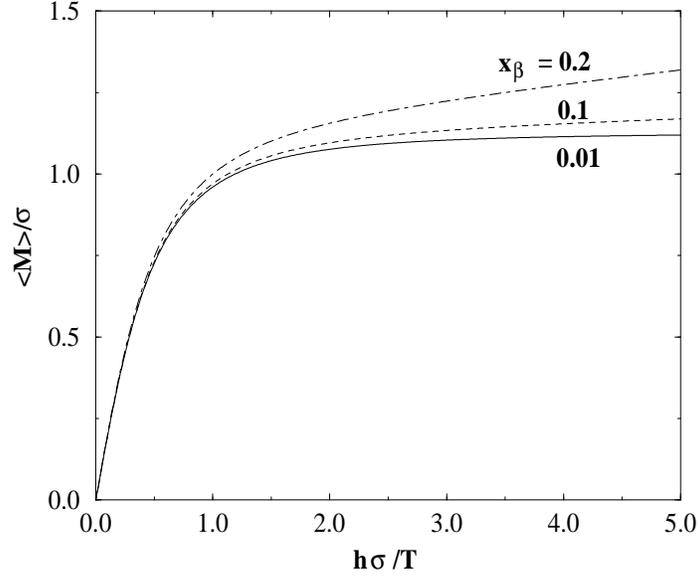,width=10cm,height=8cm}}
\caption{Variations of $\overline{M(h)}/\sigma$ versus
$h \sigma / T$ in the toy model calculation
(see (\protect\ref{eq-geom-av})), for various
values of the parameter $x_{\beta} =
\sqrt{ \overline{\sigma^2_{\beta}}}/\sigma$.
}
\label{fig-aiman}
\end{figure}

Within this toy-model, it is straightforward to 
calculate the magnetic field dependence of the
average magnetization for a fixed value
of $\Delta$. To do so, we notice that
the magnetization distribution in our toy-model
is nothing but the convolution of
$P_{\Delta,0}$ and $P_{0,\sigma_{\beta}}$.
As a consequence, 
$$
\langle \exp{(\beta h M)} \rangle_{\Delta,\sigma_{\beta}}
= \langle \exp{(\beta h M)} \rangle_{\Delta,0}
\langle \exp{(\beta h M)} \rangle_{0,\sigma_{\beta}}
,
$$
from what we deduce the average magnetization for a given
value of $\Delta$:
$$
M(h) = \sigma_{\beta}^2 \frac{h}{T} +\Delta \tanh{
\left(\frac{ h \Delta}{T} \right)}
.
$$
For a fixed $\Delta$, and for a magnetic field smaller than
the cross-over magnetic field $h^*$ defined as
\begin{equation}
\label{eq-hstar}
\left( \frac{h^*}{T} \right)^2 = 
\frac{3 (\sigma_{\beta}^2 + \Delta^2 )}{\Delta^4}
,
\end{equation}
the magnetization response is linear, whereas it is non linear
for magnetic fields stronger than $h^*$.
Since $\Delta$ and $\sigma_{\beta}$ scale like
$\sqrt{N}$, the cross-over field $h^*$ is small even
for large systems. More precisely, we now average the
magnetization over the geometry:
\begin{equation}
\label{eq-geom-av}
\overline{M(h)} =
\overline{\sigma_{\beta}^2} \frac{h}{T} +
\int_{- \infty}^{+ \infty}
P( \Delta) \Delta \tanh{\left( \frac{h \Delta}{T} \right)}
d \Delta
= 
\overline{\sigma_{\beta}^2} \frac{h}{T} +
\frac{\sigma}{2 \sqrt{\pi}} \int_{- \infty}^{+ \infty}
u \tanh{\left(\frac{h \sigma }{T}u\right)}
\exp{(- u^2 / 4)} du
.
\end{equation}
We should distinguish between the two regimes
\begin{center}
\begin{tabular}{ll}
{ Weak fields: } $h \ll T/\sigma$
& $M(h) \simeq (2 \sigma^2  +
\overline{\sigma_{\beta}^2}) h / T $\\
{ Intermediate fields: } $T / \sigma \ll h \ll T \sigma /
\overline{\sigma_{\beta}^2}$
&  $M(h) \simeq 2 \sigma / \sqrt{\pi} + 
\overline{\sigma_{\beta}^2} h/T$
.
\end{tabular}
\end{center}
As the magnetic field increased from zero, the response
to the external field is first linear, and, for magnetic
fields of the order of $T/\sigma$, cross-overs to 
a non linear behavior. This behavior is shown in
Fig. \ref{fig-aiman} for various values of the
ratio $x_{\beta} = \sqrt{ \overline{\sigma^2_{\beta}}}/\sigma$.

\subsection{Comparison with simulations}
\label{sec-comparison}
\begin{figure}
\centerline{\psfig{file=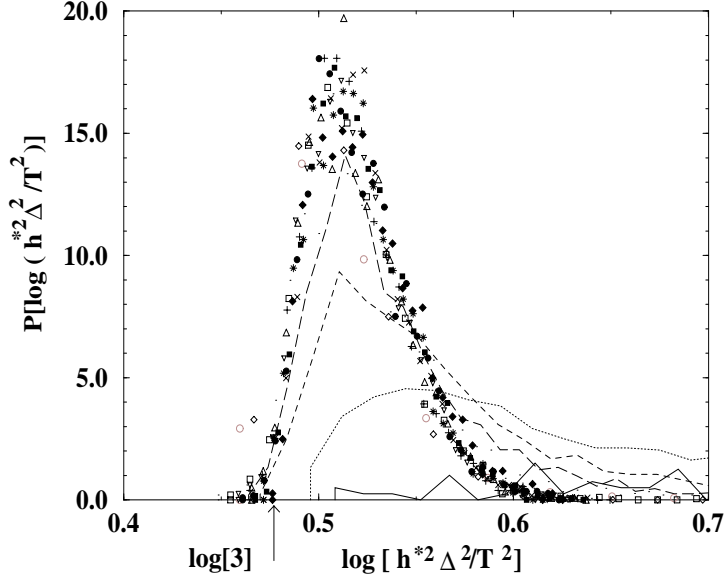,width=10cm,height=8cm}}
\caption{Histogram of $\log{((h^*)^2 \Delta^2 / T^2)}$, for
different values of the N\'eel state magnetization
$\Delta$. The lines correspond to
$\Delta = 1, 2, 3, 4$ and the symbols
to values of $\Delta$ between $5$ and $19$.
The temperature is $T=0.4$. $2000$ clusters
contained in the 20$\times$20 square
were generated for each value of $\Delta$.
}
\label{fig-rescale}
\end{figure}

We now compare the predictions of our toy-model for
the equilibrium magnetic response of percolation
clusters to numerical calculations.
We have generated 2000 clusters
for each value of the N\'eel state magnetization 
$\Delta$. All these clusters are contained
inside the 20$\times$20 square. In order to compare
with the toy-model results (\ref{eq-hstar}),
we have calculated for each of these clusters
the cross-over field $h^*$ defined by the
equality of the linear and cubic terms
in the cumulant expansion (\ref{eq-expansion}):
\begin{equation}
\left( \frac{h^*}{T} \right)^2 =
\left|
\frac{ 6 \langle M^2 \rangle_0 }
{ \langle M^4 \rangle_0 - 3 \langle M^2 \rangle_0^2} \right|
,
\end{equation}
the magnetization distribution in a zero magnetic
field being calculated with the Swendsen-Wang
algorithm \cite{Swendsen}.
We have shown
in Fig. \ref{fig-rescale} the histogram
of $\log{((h^*)^2 \Delta^2 / T^2)}$.
In the regime $\Delta \gg \sigma_{\beta}$,
Eq. (\ref{eq-hstar}) becomes $(h^*/T)^2 = 3 / \Delta^2$:
the histograms in Fig. \ref{fig-rescale} would be
a $\delta$ function located on the value
$\log{3}$. Even if the histograms in
Fig. \ref{fig-rescale} have a finite width, we see that
the scaling $(h^*/T)^2 \sim 1 / \Delta^2$ still holds,
at least for not too small values of $\Delta$.
This suggests that $\sigma_{\beta}$ is not $\Delta$-independent
but is rather distributed, and scales on average like $\Delta$.
We attribute this behavior to the existence of low
energy droplet excitations. Since these excitations
correspond to large magnetic domains, the magnetization
induced by reversing these domains should be coupled to
the total N\'eel state magnetization $\Delta$.

\section{Thermoremanent magnetization experiment}
\label{sec-thermoremanent}
In this section, we recall how the function
$X(C)$ characterizing the FDT violation
can be obtained from the magnetization relaxation
in the thermoremanent magnetization experiment. We first assume
linear response.
In the presence of
a time-dependent magnetic field $h(t)$, the linear
contribution to the magnetization is
\begin{equation}
\label{eq-mag}
m[h](t) = \int_{- \infty}^t R(t,t') h(t') d t'
,
\end{equation}
with $R$ the kernel response.
We take here as a working hypothesis that
\begin{equation}
\label{eq-R}
R(t,t') = \beta X(C(t,t')) \theta(t-t')
\frac{\partial C(t,t')} {\partial t'},
\end{equation}
a form of the response kernel suggested by
the mean field model studies
\cite{mean_field,mean_field_2,mean_field_3}.
The auto-correlation
$C$ is
\begin{equation}
C(t_w + \tau, t_w)
= \overline{ \langle \frac{1}{N} 
\sum_{i,j} \sigma_i(t_w)
\sigma_j(t_w + \tau) \rangle}
\label{eq-autocorr}
.
\end{equation}
In most of the spin-glass models (for instance:
Sherrington-Kirkpatrick model, Edwards-Anderson
model),
the correlation length is vanishing,
the reponse is purely local in space, and the
terms $i \ne j$ vanish in (\ref{eq-autocorr}).
In the presence of a finite correlation length,
the thermoremanent magnetization is conjugate
to the {\it spatially non local} autocorrelation
(\ref{eq-autocorr}). From this point of view,
(\ref{eq-mag}),(\ref{eq-R}),(\ref{eq-autocorr})
can be safely taken as a extension of $X(C)$
to our problem, in the sense that
(i) lienar FDT reads $X(C)=1$
(ii) we recover the usual definition of $X(C)$
in the limit of a zero correlation length.

Barrat \cite{Barrat} used recently an interesting
different method to handle spatially
non local responses and non linearities:
he measured the staggered magnetic response to a
random field with a zero mean. In this way, the
magnetic response to the random field
is linear as a function of the width of the random
field distribution, and conjugate to the {\it local}
autocorrelations. We underline that Barrat does
not consider the same conjugate quantities
as ours, and the functions $X(C)$ are thus
different. In particular, we cannot handle
symmetry breaking within our framework. However,
in the case of the present problem of magnetic
systems on percolation clusters, the non
linearities are quite weak and, as we
will see, we can characterize their effects
on dynamics, which could not be possible
in the framework of Barrat calculations.

In the thermoremanent magnetization experiment, $h(t)$
is a step function $h(t) = h \theta(t_w-t)$, so that
the thermoremanent magnetization reads
$$
m(t_w+\tau,t_w) = \beta h
\int_0^{C(t_w+\tau,t_w)} X(q) dq
,
$$
where we have assumed $C(t_w+\tau,0)=0$ [we have indeed
checked that this quantity was vanishing in our simulations].
The function $X(C)$ is then obtained by differentiating
the magnetic response
$$
\chi(t_w+\tau,t_w) = \frac{T}{h} m(t_w+\tau,t_w)
$$
with respect to the autocorrelation:
$X(C) = d \chi / d C$.
If the waiting time
is large enough so that equilibrium has been reached,
the magnetic response $\chi(\tau)$ is $t_w$-independent,
$X(C)$ is unity, and we recover the linear FDT:
\begin{equation}
\chi(\tau) = \frac{T}{h} m(\tau) = C_{eq}(\tau)
.
\label{eq-FDT-Intro}
\end{equation}
Quite a lot of efforts have been devoted recently
to characterize how the FDT is violated in an out-of-equilibrium
situation. Analytical solutions were obtained in
the framework of mean-field models
\cite{mean_field,mean_field_2,mean_field_3}.
The fluctuation-dissipation ratio was also obtained
in numerical simulations in various models.
For instance, in the case of spin glasses,
Franz and Rieger \cite{Franz-Rieger} have analyzed the
Edwards-Anderson model in three dimensions;
more recently, Marinari {\it et al.} \cite{Marinari-et-al}
have studied the FDT violation in three and four dimensional
gaussian Ising spin glasses, and shown that the
fluctuation-dissipation ration $X(C)$ is, in these models,
equal to the static Parisi function $x(C)$.
A model of fragile glass was also studied recently \cite{Parisi}.

As explained in section \ref{sec-equilibrium},
the magnetic response
of percolating dilute antiferromagnets is not linear
for magnetic fields of the order of $T/\sqrt{N}$. 
In the absence of linear response, the equilibrium magnetization
can be expanded in powers of the magnetic field $h$,
the coefficients of this expansion being the cumulants
of the magnetization distribution (see
Eq. (\ref{eq-expansion})).
Following the
work of Gallavotti and Cohen \cite{Gallavotti-Cohen},
Kurchan \cite{Kurchan} recently proposed an extension of
the FDT to incoporate the effects of non linear response.
However, we cannot use here this generalization in
our Monte Carlo simulations since this would
involve the calculation of the time-dependent magnetization
distribution, included the tail where the magnetization
is opposite to the magnetic field.
For our purpose, we take here
as a working phenomenological hypothesis that the
thermoremanent magnetization is given by (\ref{eq-mag}), with
the magnetic-field dependent response kernel
$$
R_h(t,t') = \beta X_h(C(t,t')) \theta(t-t')
\frac{\partial C(t,t')} {\partial t'}
.
$$
In the presence of non linearities and out-of-equilibrium
dynamics, $X_h(C)$ contains contributions both from the
non linearities and the aging dynamics. However,
in the limit of large waiting times, the system has
equilibrated and thus only the non linearities contribute
to $X_h(C)$. In the opposite limit of small waiting
times, the non linearities do not contribute
to $X(C)$, and, in this limit, a contact
can be made with FDT violations in other systems,
especially euclidian coarsening dynamics \cite{Barrat}.

\section{Numerical results for $X(C)$}
\label{sec-results}
\begin{figure}
\centerline{\psfig{file=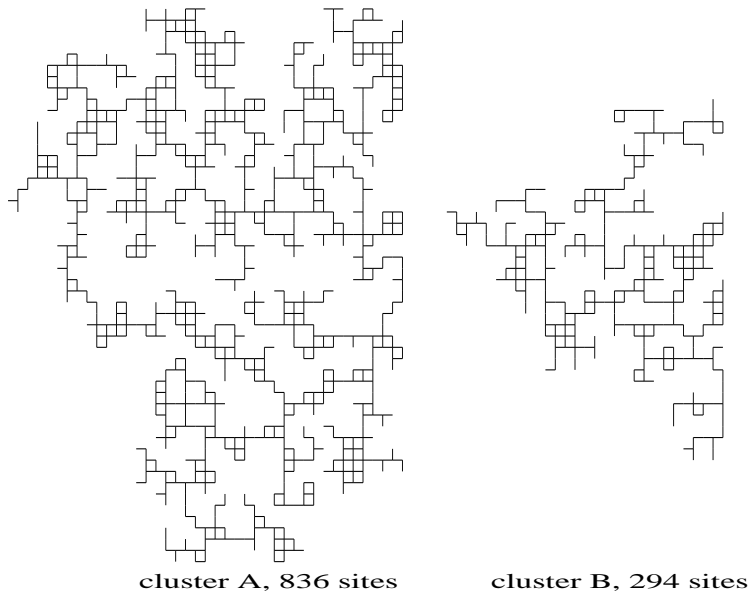,width=10cm,height=8cm}}
\caption{The two clusters that we studied. The cluster
A contains $N=836$ sites, and $\Delta=36$. The cluster
B contains $N=294$ sites, and $\Delta=16$.
}
\label{fig-dessins}
\end{figure}

We now present our numerical calculations of $X(C)$.
Our simulations were carried out on two clusters:
a cluster $A$ with $N=836$ sites, 
$\Delta=36$, and a smaller cluster
$B$ with $N=294$ sites and $\Delta=16$.
These two clusters are shown in Fig. \ref{fig-dessins}.

We first present in section \ref{sec-eq-dyn}
the equilibrium dynamics: the waiting time is long
enough for the system to have equilibrated in
the external magnetic field, and, on the basis
of the arguments presented in section
\ref{sec-equilibrium}, we expect sensible
non linear effects. 
In fact, the relaxation time is finite
even in the thermodynamic limit (interrupted aging).
As the size of the system increases,
the relaxation time will first increase, due to zero energy
barriers scaling like $J \ln{N}$ \cite{Rammal-Benoit},
and saturate when the linear size $N^{1/\overline{d}}$
becomes larger than the correlation length given by \cite{Coniglio}
$$
\xi_T \sim \exp{\left(\frac{2 J \nu_P}{T} \right)}
,
$$
with $\nu_P$ the percolation critical exponent.
The limit of small waiting times is next presented
in section \ref{sec-out-eq-dyn}. In this
situation, the magnetic response is linear.
The combined effects of non linearities and
out-of-equilibrium response arising for intermediate
waiting times are next presented in section
\ref{sec-intermediate}.
Finally, the dependence on $\tau$ of the autocorrelation
and the magnetic response are presented in section
\ref{sec-tau-dependence}.

\subsection{Large waiting times: equilibrium dynamics and
non linear response}
\label{sec-eq-dyn}
\begin{figure}
\centerline{\psfig{file=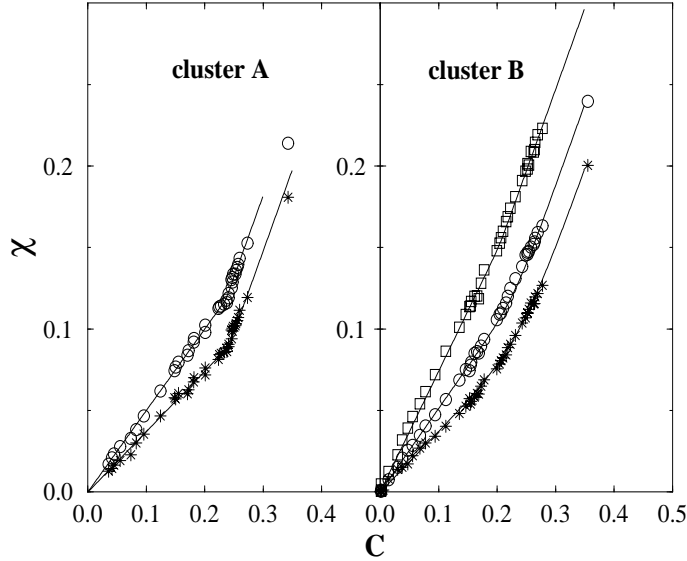,width=10cm,height=8cm}}
\caption{Variation of the magnetic response
$\chi_h(C)$ versus the autocorrelation $C$ in
equilibrium relaxation.
The temperature is $T=0.8$, and the waiting
time is $t_w = 10^5$.
The magnetic fields are
$h=0.1$ (squares), $h=0.2$ (circles) and
$h=0.3$ (crosses). The curves have been fitted
along the procedure described in the text.
}
\label{Fig4}
\end{figure}

\begin{figure}
\centerline{\psfig{file=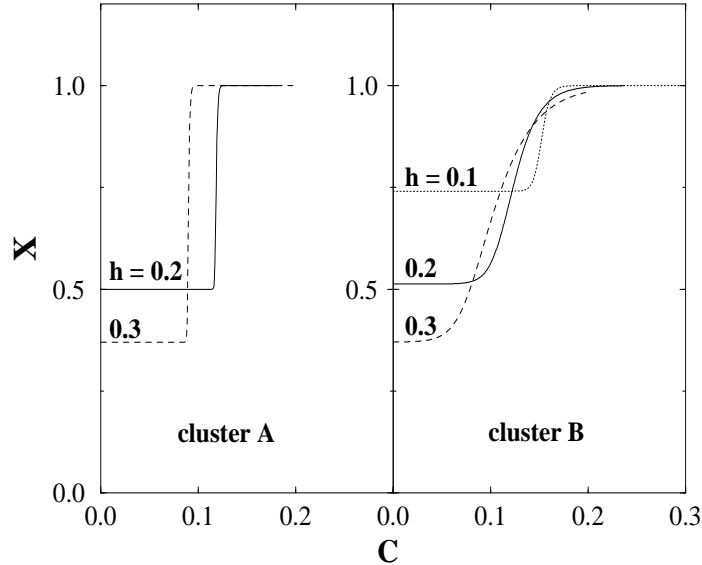,width=10cm,height=8cm}}
\caption{Variation of $X_h(C)$ versus $C$ in
equilibrium relaxation. The temperature is $T=0.8$. 
The variations of $X_h(C)$ are deduced from the
simulations presented in  Fig. \protect \ref{Fig4}.
}
\label{Fig5}
\end{figure}

We first examine the regime of a large waiting time $t_w$,
large enough for the magnetic response to be independent on
$t_w$.
In practice, we systematically checked that the
magnetic response was unchanged when the
waiting time was increased by a factor of $10$.
The magnetic response $\chi_h$ is plotted as a function
of the autocorrelation $C$ in Fig. \ref{Fig4}
for the clusters A and B.
We clearly observe on
Fig. \ref{Fig4} important non linear effects
since the magnetic response $\chi_h$ depends
explicitly on the magnetic field $h$, even
at the relatively high temperature $T=0.8$.
In the short time limit, we observe a behavior
of the type $\chi_h(C) = C - C^{(0)}_h$, whereas in
the long time limit, $\chi_h(C) = X^{(0)}_h C$.
In order to interpolate between these two behaviors,
we have fitted our numerical results to the form
\begin{equation}
\label{eq-fit-X}
X_h(C) = \frac{d \chi_h(C)}{d C}
= (1-X^{(0)}_h) f_{C^*_h,\lambda_h}(C) + X^{(0)}_h
,
\end{equation}
with
$$
f_{C^*_h,\lambda_h}(C) = 
\left( 1 + \exp{\left( - \frac{C - C^*_h}{\lambda_h}
\right)} \right)^{-1}
,
$$
where $\lambda_h$ controls the width of the cross-over
between the short and the long time regimes, and
$C^*_h = C^{(0)}_h / (1 - X^{(0)}_h)$. The fits
obtained in this way
are shown in Fig. \ref{Fig4}, and, once the three
parameters have been adjusted, a very good agreement
with the simulation data is obtained.
The variations of $X_h(C)$ deduced from the fits
are shown in Fig. \ref{Fig5} for the
same simulations as in Fig. \ref{Fig4}.
We observe in Fig. \ref{Fig5} that
$X_h(C)$ cross-overs from unity
at short times to a finite value $X^{(0)}_h$
in the long time relaxation. If the response
to the external magnetic field was linear,
one would expect that $X(C)=1$.
Even though we could not
address this question here, we expect
a non linear FDT of the type \cite{Kurchan} to hold
in the long waiting time limit.

\subsection{Small waiting times: out-of-equilibrium dynamics and
linear response}
\label{sec-out-eq-dyn}
\begin{figure}
\centerline{\psfig{file=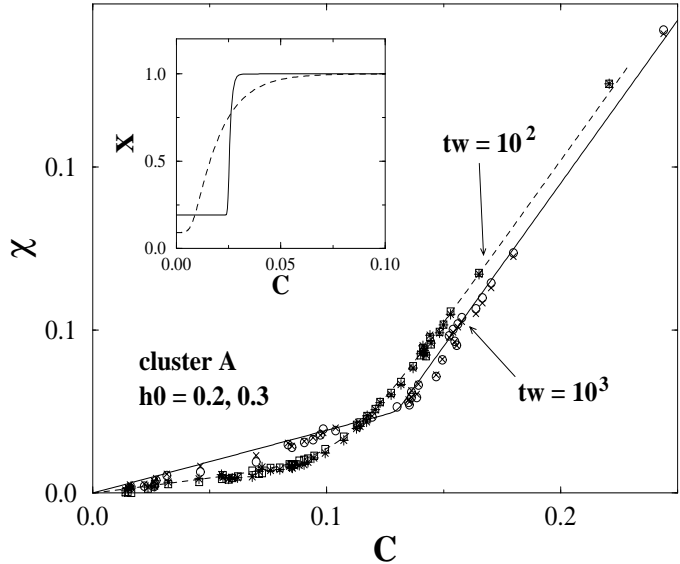,width=10cm,height=8cm}}
\caption{Variations of the magnetic response
$\chi(C)$ versus the autocorrelation $C$ in the
short waiting time limit ($t_w = 10^2, 10^3$),
and $h=0.2, 0.3$. The corresponding variations of
$X(C)$ are shown in the insert.
}
\label{Fig6}
\end{figure}

In the short waiting time limit,
the thermoremanent magnetization
is linear as a function of the magnetic field.
We have shown in
Fig. \ref{Fig6} the variations of the magnetic
response $\chi$ versus the autocorrelation $C$ for
the two values of the magnetic field $h=0.2, 0.3$,
and $t_w = 10^2, 10^3$. Linear
response is clearly observed.
The variations of $X(C)$ are shown in 
the insert of Fig. \ref{Fig6}.
Interestingly, the variations of $X(C)$ in this situation
where out-of-equilibrium effects are dominant are
qualitatively the same as the ones in section
\ref{sec-eq-dyn}: $X$ cross-overs from
unity at short times to a finite value in the 
long time limit. We have no understanding of the
reason why the variations of $X(C)$ are qualitatively
the same in the small and large waiting time limits,
where deviations from linear FDT originate respectively
from the out-of-equilibrium dynamics and non 
linear response.

The fact that $X(C)$ is finite in the aging regime
is a quite noticeable difference
with euclidian coarsening \cite{Barrat}, where $X$
is vanishing in the aging regime (the linear response
kernel $R$ in (\ref{eq-mag}) is zero in this regime). 
The long term memory of coarsening dynamics on percolating
structures is thus stronger than for euclidian dynamics,
which is maybe not surprising on the general grounds
recalled in the introduction: a ``droplet'' of size $N$
has a zero temperature energy barrier
scaling like
$\ln{N}$ \cite{Rammal-Benoit,Rammal,Henley},
and a finite energy
of the order of $2 {\cal C} J$,
${\cal C}$ being the order of
ramification \cite{LesHouches}.

\subsection{Intermediate waiting times: out-of-equilibrium
dynamics and non linear response}
\label{sec-intermediate}
\begin{figure}
\centerline{\psfig{file=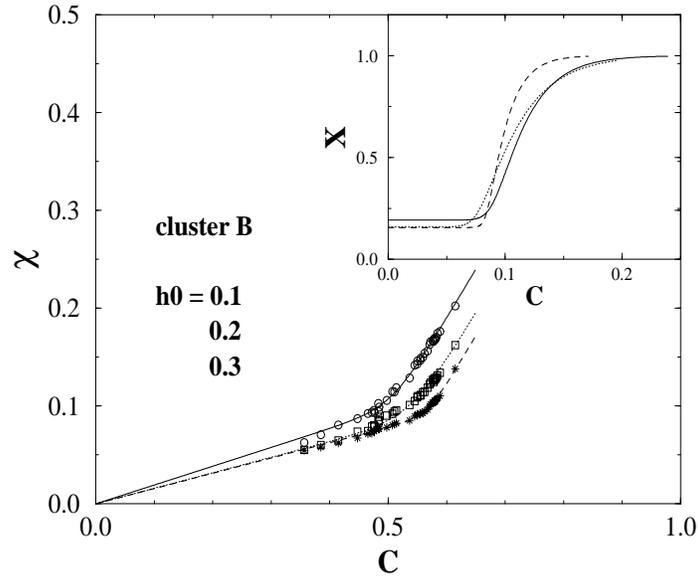,width=10cm,height=8cm}}
\caption{Variations of the magnetic response
$\chi_h(C)$
versus the autocorrelation $C$ for the cluster B, 
$T=0.55$, $t_w=10^6$, $h=0.1$ (circles),
$h=0.2$ (squares) , $0.3$ (crosses). The
insert shows the corresponding variations
of $X(C)$.
}
\label{Fig7}
\end{figure}

In order to examine the conjugate effects of nonlinearities
in the magnetization response and out-of-equilibrium dynamics,
we carried out the thermoremanent magnetization simulation
with the cluster B at the temperature
$T=0.55$, and for a waiting time $t_w = 10^6$.
The results are shown in Fig. \ref{Fig7}, with
$\tau$ up to $10^7$.
We have checked that equilibrium was not reached by
carrying a simulation with a waiting time
$t_w = 10^7$. On the
other hand, the magnetic response $\chi_h$
depends explicitly on the magnetic field $h$, as
is visible in Fig. \ref{Fig7}.
We observe that $X_h(C)$ can still be fitted by the
form (\ref{eq-fit-X}), even though we could not reach
very small values of the correlation and magnetic responses,
even for $\tau = 10^7$.

\subsection{$\tau$-dependence of $\chi(t_w+\tau,t_w)$
and $C(t_w+\tau,t_w)$}
\label{sec-tau-dependence}
\begin{figure}
\centerline{\psfig{file=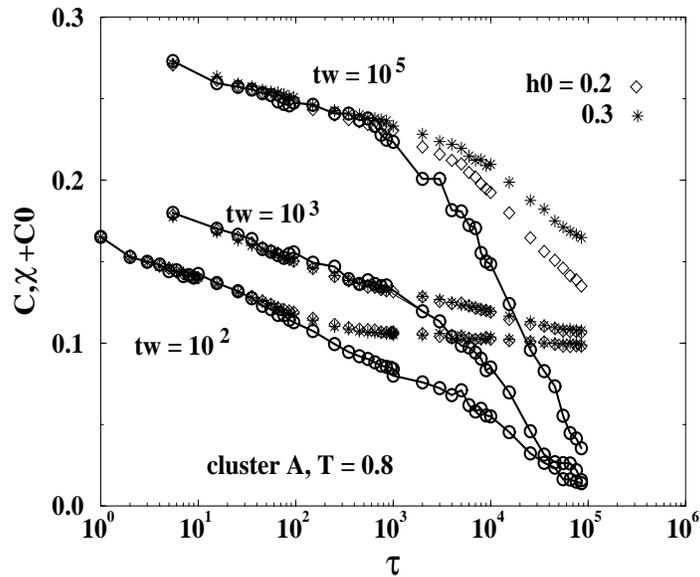,width=10cm,height=8cm}}
\caption{Variations of $\chi(t_w+\tau,t_w) +
C^{(0)}(t_w)$ (diamonds: $h=0.2$,
crosses: $h=0.3$) and $C(t_w+\tau,t_w)$ (solid lines)
versus $\tau$. The waiting times are
$t_w = 10^2, 10^3, 10^5$. The temperature
is $T=0.8$.
}
\label{Fig-tau1}
\end{figure}

\begin{figure}
\centerline{\psfig{file=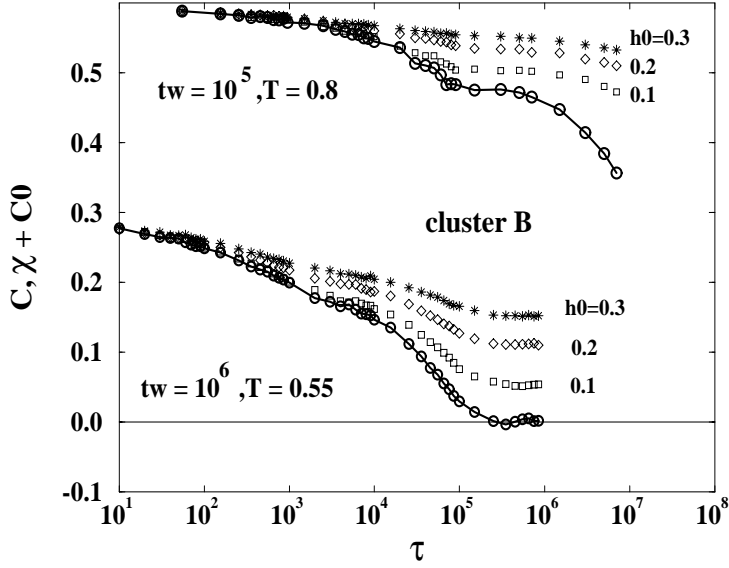,width=10cm,height=8cm}}
\caption{Variations of $\chi(t_w+\tau,t_w) +
C^{(0)}(t_w)$ (squares: $h=0.1$, diamonds: $h=0.2$,
crosses: $h=0.3$) and $C(t_w+\tau,t_w)$ (solid lines)
versus $\tau$. The waiting times are
$t_w = 10^5$ ($T=0.8$), and $t_w=10^6$ ($T=0.55$).
}
\label{Fig-tau2}
\end{figure}

In spin-glass models, the short time regime
$\chi(t_w+\tau,t_w) = C(t_w+\tau,t_w)
- C^{(0)}(t_w)$ is valid up to a time
$\tau^*$ of the order of the waiting time $t_w$
\cite{Franz-Rieger}. As shown in
Fig. \ref{Fig-tau1}, we indeed observe such
a dependence of $\tau^*$ in the out-of equilibrium
situation: $\tau^*$ is of the order
of $10^2$ for $t_w=10^2$, and of
the order of $10^3$ for $t_w=10^3$.
However, for larger waiting times,
non linearities significantly reduce $\tau^*$ 
($\tau^* \simeq 10^3$ for $t_w=10^5$
in Fig. \ref{Fig-tau1}
).
We observe in Fig. \ref{Fig-tau2} in the
case $t_w=10^5$, $T=0.8$ (equilibrium situation) that
$\tau^*$ is also a function of the magnetic field $h$
(the value of $\tau^*$ for $h=0.1$ is
one order of magnitude larger than for $h=0.2$).
This effect is also visible in the out-of-equilibrium
simulation shown in Fig. \ref{Fig-tau2} ($T=0.55$).
However, from our simulations, we cannot make a precise
statement on the variations of $\tau^*$ as a function
of $h$ for large waiting times.
\section{Conclusions}
\label{sec-conclusions}

We have thus carried out Monte Carlo simulations of
the violation of the linear FDT in dilute
percolating antiferromagnets.
We have shown that these systems exhibit non linear response
for magnetic fields of the order of $T/\sqrt{N}$.
In the small waiting time regime, the thermoremanent
magnetization is linear in the magnetic field,
but depends explicitly on the waiting time.
On the other hand, for sufficiently large waiting
times, the system has equilibrated (interrupted aging),
and the magnetic response is non linear.
Interestingly, in both situations, as well as
in the intermediate situation where
both out-of-equilibrium and non linear effects
come into account,
the function $X(C)$ characterizing the deviations from
linear FDT has qualitatively the same shape:
it is unity at short times and cross-overs to a
constant finite value for long times, the
cross-over occurring at $\tau^*$. In the
small waiting time limit, $\tau^*$
is of the order of the waiting time $t_w$.
For larger waiting times, 
non linearities strongly reduce $\tau^*$
as the magnetic field is increased.
By comparison with domain growth processes
in non diluted lattices, the aging part of 
the dynamics shows stronger long-term memory,
due to the existence of large scale
low-energy `droplet' excitations.

\medskip

{\bf Acknowledgments} R.M. thanks A. Barrat,
L.F. Cugliandolo, J. Kurchan, S. Franz and P.C.W. Holdsworth
for stimulating discussions.
Part of the calculations presented
here were carried out on the CRAY T3E computer of the Centre de
Calcul Vectoriel Grenoblois of the Commissariat \`a
l'Energie Atomique.

\end{document}